\pdfoutput=1
\documentclass[twocolumn,aip,rsi,amsmath,amssymb,reprint]{revtex4-1}
\usepackage{graphicx}
\usepackage{dcolumn}
\usepackage{bm}
\usepackage{threeparttable}
\usepackage[colorlinks = true,
            linkcolor = blue,
            urlcolor  = blue,
            citecolor = red, 
           anchorcolor = blue]
            {hyperref}
\usepackage{url}
\begin{document}
\preprint{AIP/123-QED}
\title{The staged Z-pinch as a potential high-gain fusion energy source: Rebuttal to I. R. Lindemuth et al.}
\author{E. Ruskov,}
\email{emil@miftec.com}
\author{P. Ney, H. U. Rahman  \\ 
\it Magneto-Inertial Fusion Technology Inc. Tustin, CA}

\date{\today}

\begin{abstract}
Magneto-Inertial Fusion Technology Inc.  has been working on a Z-pinch concept where a high atomic number liner is compressing a fusion fuel (deuterium-deuterium, or deuterium-tritium) target. The viability of this so called Staged Z-pinch (SZP) concept as a potential high-gain fusion energy source has been questioned in a recent publication by Lindemuth et al$^1$.  The authors attempted to reproduce previously published MACH2 simulation results$^{2-4}$ for Z-machine parameters using three different MHD codes: Hydra, Raven and MHRDR. Their  conclusion was that "there is no conceivable modification of the parameters that would lead to high-gain fusion conditions using these other codes".  Although they used well established MHD codes to check the SZP concept, and correct input current profiles, we show that their Lagrangian formalism was likely not treating the vacuum/liner boundary properly. Proper modeling using  Lagrangian, Eulerian or Adaptive Lagrangian-Eulerian (ALE) formalism indeed confirms that  fusion energy production $>$ 1 MJ can be expected without alpha heating, and significantly higher if alpha heating is included. It is shown that  magnetosonic shocks play an important role in preheating the target plasma and in piling liner mass at the liner/target interface, which substantially increases the ram pressure just before the pinch stagnation time.

\end{abstract}

\maketitle

\section{Introduction}

Z-pinches are  one of the first thermonuclear fusion energy ideas explored. The  earliest observations of deuterium-deuterium fusion neutrons from Z-pinches were reported in 1950s$^{5-8}$, but the classical magnetohydrodynamic (MHD) and magneto Rayleigh-Taylor (MRT) instabilities limited the  fusion yield. 
In the late 1970s, Amnon Fisher and collaborators at the University of California, Irvine (UCI), created the first gas puff Z-pinch using a 200-kA pulsed power generator$^9$. Use of gas mixtures  enhanced the pinch stability and increased its radiation efficiency$^{10}$. Preionization with electron beams  improved the uniformity of the initial-breakdown  in a gas puff Z-pinch$^{11}$ and increased the magnetic flux compression, allowing   amplification  of the axial magnetic field $B_z$ and stabilization of the Z-pinch$^{12,13}$.  Later on gas puff Z-pinch experiments were undertaken  on several multi megaamperes pulse power generators with a goal of increasing either the X-ray$^{14,15}$ or the neutron yield$^{16-19}$.

 The UCI experiments  led to a concept called Staged Z-pinch (SZP) where the energy to the final load is transferred in successive stages and the rate of energy transfer increases in each stage.  The SZP name was initially used$^{20}$ for an annular shell (i.e. liner) compressing  an on-axis cryogenic deuterium fiber (i.e. target): the current pre-pulse in the fiber  pre-magnetized the liner, which was then compressed with the main Z-pinch current pulse through the liner; if the azimuthal magnetic flux is  conserved inside the liner, the initial pre-pulse current can grow to a very large value on a fast time scale.  The key  SZP feature  is the control and mitigation of the magneto-Rayleigh-Taylor instability which allows  formation of a stable target plasma, even though the liner plasma becomes unstable$^{21}$. The concept further evolved$^{22}$ to include gas-puff liners. It was theorized that by using a high atomic number (i.e. high-Z) liner, the liner would radiatively cool down and thus facilitate the magnetic field diffusion.

The Z Pulsed Power Facility  at the Sandia National Laboratory is the most powerful Z-pinch machine in the world. It has  20 MJ  stored energy in large capacitor banks and can deliver up to a 26 MA load current pulse  with a 100 ns rise time$^{23}$. Megajoules of X-ray energy over a few nanosecond period were radiated in plasmas created from wire array loads. Such plasmas are of great scientific and technical interest, for example in studies related to fusion, atomic physics and  laboratory astrophysics$^{24-26}$. More recently, extensive Magnetized Liner Inertial Fusion (MagLIF) experiments are carried on this machine, where a beryllium  liner is used to compress laser preheated deuterium target plasma$^{27-29}$.  This preheating scheme introduces additional complexities$^{30}$ (for example, the uniformity of  the laser energy deposition and the propagation of the burn wave) which are under active investigation. 

Our recent experiments at the 1 MA Nevada Terawatt Facility  (University of Nevada, Reno) investigated the compression of a deuterium  target by argon (Z=18) and krypton (Z=36) liners$^{31}$.
The pinch  implosion dynamics  was studied with the radiation-MHD code MACH2$^{32}$  using initial conditions approximating the experiments. MACH2 simulations  confirmed the diffusion of the azimuthal magnetic field through the liner, indicating that the associated magnetic field pressure contributed to  the  target acceleration. Shock waves then develop in the target plasma,  preheating it to several hundred eV.  Finally, the target is adiabatically compressed to stagnation, reaching volume-averaged ion temperatures, for the krypton liner case, of 4 keV.   Neutron yields of up to $2\times10^9$ were measured in argon, and up to $2.5\times10^{10}$  in krypton  liner experiments, and they were in reasonable agreement with the MACH2 predictions.

The shock aided preheating in the Staged Z-pinch has inherent simplicity, and the  preheating strength is largely determined by the implosion velocity and the atomic composition of the liner. 
In a recent paper by  Lindemuth et al$^1$.  the viability of  the SZP concept   as a  potential high gain fusion energy source was questioned.  The authors attempted to reproduce our  published MACH2  results for the Z-machine by using three different MHD codes (Hydra, Raven and MHRDR) running in 1-D Lagrangian mode, and they concluded that "there is no conceivable modification of the parameters that would lead to high-gain fusion conditions using these or any other codes". They also recommended  that  SZP should not be considered as a potential high-gain fusion energy source.  Our papers were identified as  SZP1$^2$ (xenon liner), SZP2$^3$ (silver liner with Gaussian  radial mass-density profile) and SZP3$^4$ (silver liner with flat radial mass-density profile); they all modeled compression of a 50-50 \% deuterium-tritium target.  For the sake of clarity and brevity,  in this paper we discuss only  the SZP2 case. We tested the other two cases, SZP1 and SZP3, and confirmed that they produce high fusion gain as well.

In section II we reproduce the  SZP2 results of Lindemuth et al. by running the MACH2 code in 1-D Lagrangian mode and by using their  computational grid: one liner block  and one target block, with 64 grid points in each block. This verifies the code as a SZP  modeling tool. We further show that this set-up is not appropriate as it leads to too few computational cells in the outermost liner region, and an incorrect calculation of the azimuthal magnetic field $B_\theta$. When a dedicated, third vacuum block is added to the simulation, the problems with the $B_\theta$ profile calculation are alleviated and the plasma current calculated from Ampere's law at the liner/vacuum boundary is identical to the current calculated from circuit equations. We want to point out that in most of our simulations the current is derived from circuit equations, with dynamically calculated pinch inductance and resistance at each time step. This current in essence drives the simulation, which next calculates the $B_\theta$ magnetic filed, the induced current in the plasma,  the associated ohmic heating, and so on. The three block Lagrangian simulation predicts  much higher fusion energy yield, especially when $\alpha$-particle heating is included. We  confirmed these results by running 1-D pure Eulerian and ALE simulations. 

In section III we discuss the important role of shocks in plasma preheating, and in concentrating mass at the liner/target boundary which enables significant final adiabatic target heating.  Improperly calculated azimuthal magnetic field $B_\theta$ results in weaker shocks, less mass concentration and weaker ram pressure in the last nanosecond of the target compression  which limits its temperature growth to 1 keV.

More realistic, two dimensional simulation results based on high resolution pure Eulerian and ALE models are presented in section IV. The role of $\alpha$-particle heating and plasma radiation are  discussed at length in section V.

\section{ Staged Z-pinch 1-D Lagrangian MACH2 simulations: Code Verification}

Any code attempting to simulate  a real  physical system has to undergo  verification and validation.  Code verification confirms that the computer model describing the system is correctly implemented, while code validation compares the simulation results with measurements and gives credence to the code usefulness for analyzing the physical system.

The Multi-block Arbitrary Coordinate Hydromagnetic (MACH2) code$^{32}$  is a multi-material, single fluid, three temperature resistive MHD code, developed by the Center for Plasma Theory and Computation at the Air Force Research Laboratory, Phillips Research Site. It solves for the  following set of equations: mass continuity, single fluid momentum, electron and ion specific internal energy,  radiation energy  and the Faraday's law  for the magnetic field. The radiative losses are calculated with a single-group, flux limited, non-equilibrium radiation model. Plasma equations of state and transport variables (radiation opacities, thermal and electrical conductivities, etc.) are obtained from the LANL SESAME tables. 

This  2 1/2 dimensional code has  an adaptive mesh generator which can adjust the computational grid according to user specified criteria. Its  Arbitrary Lagrangian-Eulerian (ALE) implementation allows simulations to be run in pure Lagrangian, pure Eulerian, pure Eulerian axially and Lagrangian radially, or any other  combination. The adaptive algorithm can adjust the grid spacing, depending on the magnetic field or plasma pressure gradients (or both), providing increased computational accuracy  in regions with  large spatial changes of these quantities while, in principle, conserving computational time. 
The fixed Eulerian method, where the computational grid is constant for all time steps in the simulation, is the most straightforward to conceptualize and analyze.  However,  to properly resolve  important phenomena driving the system dynamics, it may require subdividing the simulation domain into numerous blocks with a sufficiently high number of cells in each block.

MACH2 has been  successfully used for studies of plasma opening switches$^{33,34}$, explosive magnetic generators, inertial-confinement fusion and alternate fusion concepts$^{35}$, compact toroid schemes$^{36,37}$,  and Z-pinches with solid liners$^{38,39}$.   Therefore,  the code itself has been extensively verified and validated. A possible question remains whether it was correctly used in  our SZP  simulations; in particular, whether adequate   boundary conditions and spatial resolution were provided. 

Limited computational resources at our disposal years ago led us to use the ALE method for our previous papers. For this paper all three major methods were used: pure Lagrangian, pure Eulerian and ALE with feedback on  magnetic field and plasma pressure gradients. Both R-space (1-D) and  RZ-space (2-D) calculations were done.

MACH2 has a self consistent circuit modeling capability, and it has been modified to accurately model the refurbished Z pulsed power machine at Sandia National Laboratory$^{40}$. The simplified R-L circuit  has the following parameters: $\rm Z_0 = 0.18~\Omega$, L = 6.64 nH, C = 8.41 nF, $\rm L_0$ = 6 nH and R = 10 $\Omega$.

Lindemuth et  al. correctly state that  "Although various test problems can be used for partial verification, at some point, only code comparisons such as reported in this paper will provide confidence that the simulations can accurately guide an experimental program". Therefore, our  first attempt was to verify the MACH2 model  of the Staged Z-pinch by using  exactly the same parameters  as those used in  the Hydra, Raven and MHRDR models; computational grid with only one liner and one target block, with 64 grid points in each of them; initial target density of $\rm 9.8\times10^{-3}$ gm$\rm/cm^3$ from 0 to 2 mm;   initial silver liner density of 0.6 gm/$\rm cm^3$ with uniform distribution from 2 mm to 3 mm;  and  initial temperature of 2.0 eV  applied to both regions. The silver liner opacities are not publicly available, so we and Lindemuth et  al. use  opacities for dysprosium (SESMAE material 212).

Figure 1 illustrates the three distinct SZP regions (target, radiative liner  and vacuum), the plasma current flowing predominantly at the inner liner/vacuum boundary, the azimuthal magnetic field  created,  and the Lorentz force  $\bf J \times B$ which compresses the pinch. 

\begin{figure}
\centering
\includegraphics[width=3.in, keepaspectratio=true]{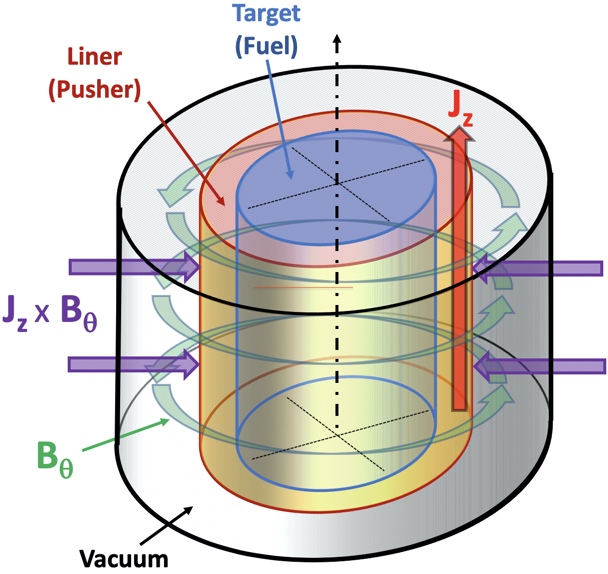}
\caption{Schematic of the staged Z-pinch showing cylindrical fuel plasma column surrounded by a high-Z liner plasma.The self-generated azimuthal magnetic field $B_\theta$ interacts with the axial plasma current $J_z$ (flowing through the liner periphery) and compresses the fuel plasma to thermonuclear conditions.}
\label{fig:1}
\end{figure}

We studied two 1-D pure Lagrangian models driven by current calculated from the circuit equation, and compared their results with the Raven code:  one with two blocks,  each of them with 64 cells, and another, which is extension of the first, where a  third eight cell vacuum block  extending from 3.0mm to 3.1mm was added. Vacuum in the three block model  is defined as a very low density  plasma region, with high resistivity, so that the  magnetic flux can  rapidly diffuse  through it.  In figures 2, 3, 4 and 5 
we refer to these models as \it Lag\_2blk \rm and \it Lag\_3blk\rm, and use the $\it alpha$ suffix if $\alpha$-particle heating was calculated.

We also run a third  model driven by   current from the Lindemuth et  al. paper. Note that we digitized the current  from that publication, which in turn  must have been digitized from our publication$^3$. The model has three blocks; a 225 cell liner block,  a 100 cell target block, and a 30 cell vacuum block, where the outer vacuum boundary is fixed at 3.1cm. In pure Lagrangian simulations the computational grid moves with the plasma fluid which preserves  high resolution in each block as the plasma profiles steepen. If the outer vacuum boundary is fixed, the pinch compression leaves wider and wider vacuum regions behind; having a fixed number of grid points, the spatial resolution in this region becomes  coarse, resulting in a crude calculation of the magnetic field.  We refer to this simulation as  \it Lag\_3blk\_hydra, \rm because that was apparently the model used in the 1-D Hydra code simulation$^{41}$.

Currents from this Lagrangian study are compared in Fig.2. The two and three block pure Lagrangian currents are essentially indistinguishable until the stagnation time (125.87 ns for \it Lag\_2blk \rm and 127.01 ns for \it Lag\_3blk\rm). Subsequently, they start deviating and a particularly steep rise is seen in the \it Lag\_3blk\_alpha \rm  case. The post-stagnation current evolution in  the two block models (with and without alpha heating) is indistinguishable.

\begin{figure}
\includegraphics[width=3.3in,keepaspectratio=true]{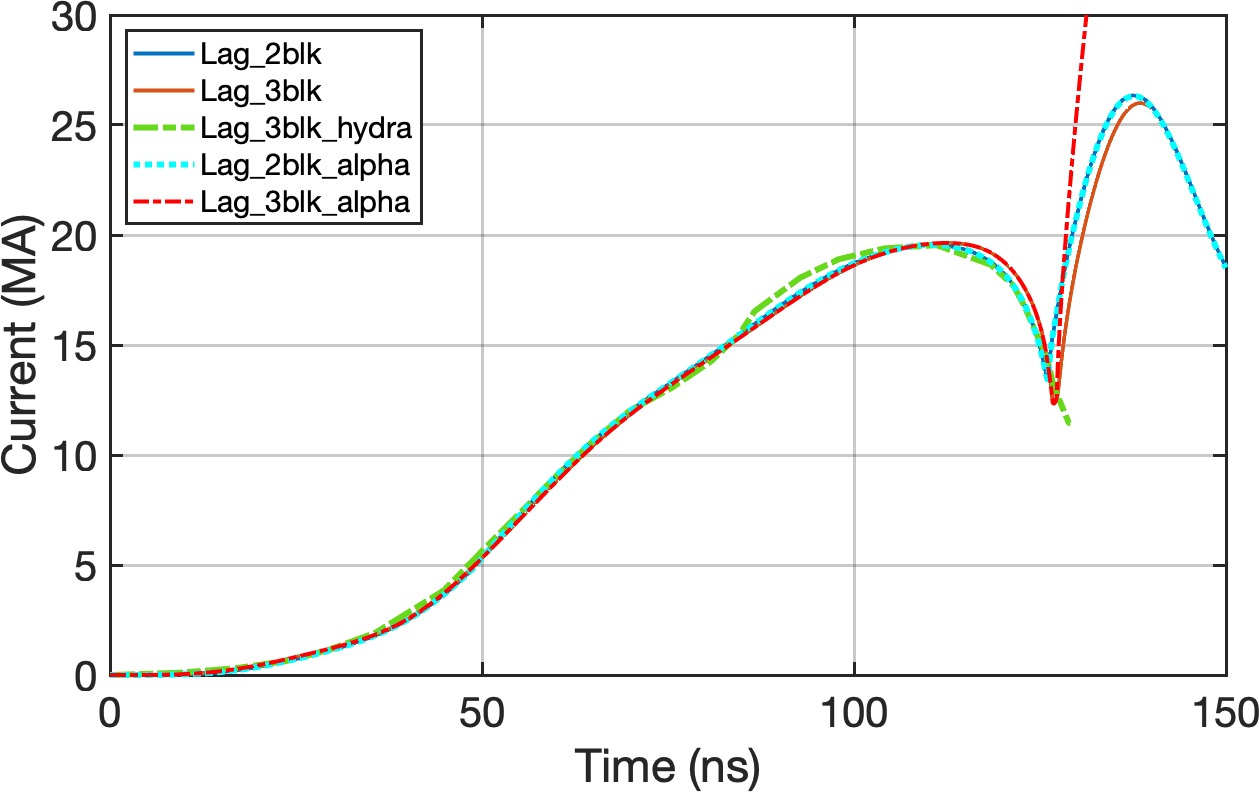}
\caption{Current profiles calculated from circuit equations with the MACH2 code, for two and three block 1-D Lagrangian simulations, with and without alpha particle heating. The Lagrangian simulation with a \it hydra \rm suffix is using three blocks, fixed vacuum boundary and input current profile from  the Lindemuth et al. publication$^1$.}
\label{fig:2}
\end{figure}

In Fig.3 the time evolution of:  (a)  interface radii, (b) interface velocities, and (c) mass averaged fuel temperatures are compared with the corresponding waveforms from the Raven code. We chose the Raven code curves because they were easier to distinguish in the Lindemuth et al. paper; the curves for the other two codes were  similar but were often obscured with identification letters and thus more difficult to digitize.  

The radial position and velocity of the liner/target interface is quite similar for both models and codes.  For the two block Lagrangian model, the  mass averaged fuel  temperature peaks at 1 keV which is the same value as in the Raven case; addition of $\alpha$-particle heating in the model does not change the maximum temperature.  For the  three block  Lagrangian model, the temperature continues rising to 2.8 keV and dramatically increases to 60 keV when $\alpha$-particle heating is included, as  highlighted in 3(d). The  mass averaged fuel temperate  the \it Lag\_3blk\_hydra  \rm model also peaks at 1 keV (not shown).

\begin{figure}
\centering
\includegraphics[width=3.3in,keepaspectratio=true]{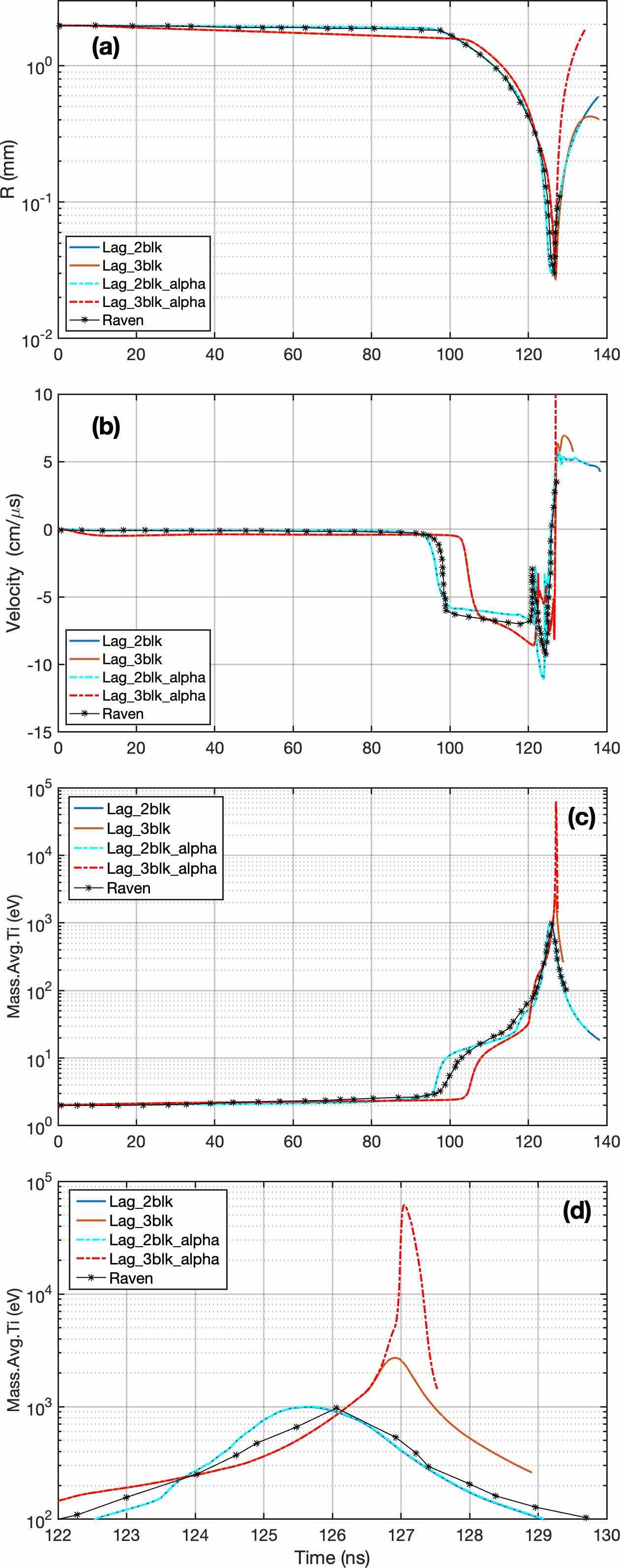}
\caption{Comparison of  two and three block 1-D MACH2 pure Lagrangian simulations, with and without alpha heating, with results from the Raven code:  (a) radius and (b) implosion velocity, both calculated at the liner/target interface, and (c)  mass-averaged fuel temperature. The crucial temperature difference among these simulations is highlighted in (d). }
\label{fig:3}
\end{figure}

The pinch compression dynamics is illustrated in Fig.4. The two block Lagrangian model has very similar  compression trajectory to the Raven case, with peak compression ratio $CR_{max}\sim64$. The three block Lagrangian models exhibit  significant shock preheating from 30 eV to 160 eV  over just 2.7 ns (119.6-122.3 ns) when the corresponding  compression ratio increases from 4 to 6.7; their peak mass averaged temperatures are  2.8 keV and 60 keV, depending whether $\alpha$-particle heating is included or not.

\begin{figure}
\centering
\includegraphics[width=3.4in, keepaspectratio=true]{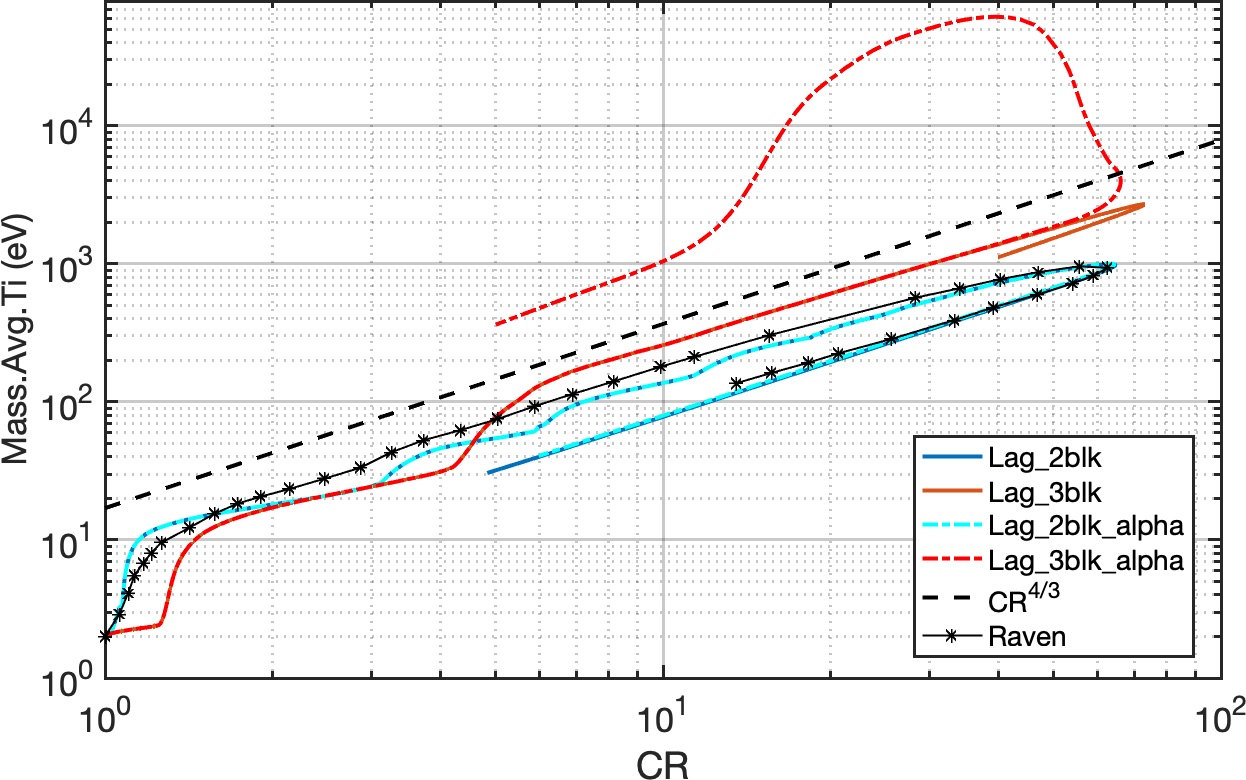}
\caption{ Comparison of  the mass-averaged fuel temperature  vs. compression ratio for two and three block 1-D MACH2 pure Lagrangian simulations, with and without alpha heating, with results from the Raven code. The curve sections that raise faster than the $\rm CR^{4/3}$ line indicate shock heating. }
\label{fig:4}
\end{figure}

What is the reason for such significant difference between the two and three block pure Lagrangian simulations? Ampere's law states that at all times the current  driving the pinch should be: $I(t)=5.0\times R(t) \times B_{\theta}(t)$, where $ B_{\theta}$ is the azimuthal magnetic field in Gauss, calculated at the outer liner radius R, which is in cm. This current  should be identical to the current driving the simulation, regardless of whether it is calculated from circuit equations or provided as direct input. Figure 5 shows that this is the case with the three block, but not with the two block pure Lagrangian simulation, where the peak current calculated from Ampere's law is 8.2\% lower. At first glance, it is perplexing why such marginally lower current leads to clamping the mass averaged target temperature to 1 keV (Fig.3d). This question will be explored in the next section where we discuss how the underlying differences in the calculated azimuthal magnetic field lead to quite different adiabatic compression strength near the stagnation time when the compression stops and the pinch starts to expand.

There is larger difference between the current driving the  \it Lag\_3blk\_hydra  \rm model and the  current derived from Amperes law; their peak values differ by 20.6\%. In order to match the pinch implosion time in the SZP2 paper, Lindemuth et al. had to lower the liner mass density by 30\% as compared to the one used in SZP2. The apparent problem with  the magnetic field calculation is the likely explanation why they had taken this course. 

In summary, by reproducing the Lindemuth et al. results (Figs.3,4) with the two block pure Lagrangian model,  we verified the MACH2 code.

\begin{figure}
\includegraphics[width=3.3in, keepaspectratio=true]{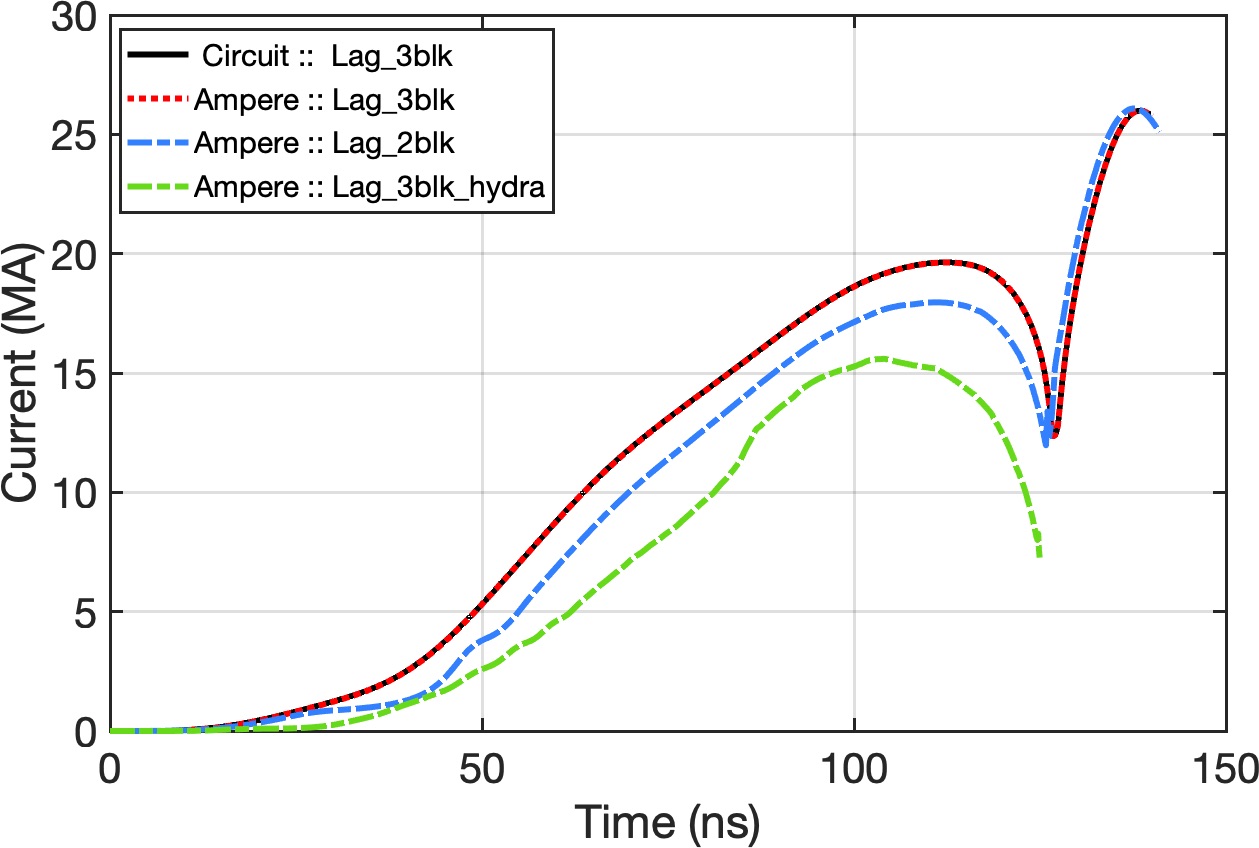}
\caption{Comparison of the current profile (black) obtained from the circuit equation, with  current profiles  calculated from Ampere's Law, for  1-D MACH2 Lagrangian simulations with for 3 blocks (red dots),  2 blocks (blue), and three blocks with fixed vacuum boundary and input current profile from  the Lindemuth et al. publication$^1$ (green).  }
\label{fig:5}
\end{figure}

\section{ Staged Z-pinch 1-D MACH2 simulations:  Fusion Yield, and the Role of shocks and ram pressure}

In addition to the Lagrangian models discussed in the previous section,  we studied ALE and  Eulerian models with various grid resolutions. The 2-D ALE and  Eulerian  models will be discussed in the next section; here we compare the corresponding  1-D  models with the Lagrangian models.

The Eulerian models were defined over  five radial blocks: three for  fuel plasma (128 cells for 0.0-0.2 mm, $\Delta R=1.6\mu m$;  128 cells for 0.2-1.0 mm, $\Delta R=6.25\mu m$; 64 cells for 1.0-2.0 mm,  $\Delta R=15.6\mu m$), one for liner plasma (64 cells for 2.0-3.0 mm, $\Delta R=15.6\mu m$), and  one for low density vacuum region (16 cells for 3.0-3.1 mm, $\Delta R=6.25\mu m$). 

The ALE models were also defined over  five radial blocks:  three for  fuel plasma (64 cells for 0.0-0.5mm, $\Delta R=7.8\mu m$; 64 cells for 0.5-1.0mm, $\Delta R=7.8\mu m$;  64 cells for 1.0-2.0mm, $\Delta R=15.6\mu m$, one for liner plasma (64 cells for 2.0-3.0 mm, $\Delta R=15.6\mu m$),  and one for low density vacuum region (8 cells for 3.0-3.1 mm, $\Delta R=12.5\mu m$ ). 
They used an adaptive mesh generator, with feedback on both the magnetic field and the plasma pressure spatial gradients, which enabled proper MHD calculations when these quantities rapidly change in a radial grid that was 2-3 times coarser  than  the Eulerian grid.  This leads to a similar spread in the  inflection times of the fusion energy curves when they reach $>$90\% of their final values.
   
\begin{table}
\caption{Stagnation time and fusion energy produced} 
\label{table}
\centering
\def\arraystretch{1.5}
\vspace{3mm}
\hspace{8mm}\begin{tabular}{|c |c |c |}    \hline
Model  & ~~ $\rm T_{stag}$ (ns) ~~& ~~ $\rm E_{fusion}$ (MJ)   ~~\\ \hline \hline

EUL\_1D               & 125.88   &  4.24      \\ \hline 
ALE\_1D               & 125.53   &   3.56     \\ \hline 
Lag\_2blk              & 125.87   &   0.03     \\ \hline 
Lag\_3blk              & 127.01   &   4.67     \\ \hline 
Lag\_3blk\_hydra  & 125.52   &   0.09     \\ \hline \hline

EUL\_1D\_alpha           & 125.79   &  196      \\ \hline 
ALE\_1D\_alpha           & 125.47   &   142     \\ \hline 
Lag\_2blk\_alpha          & 125.87   &   0.03    \\ \hline 
Lag\_3blk\_alpha          & 126.98   &   190     \\ \hline 

\end{tabular}
\end{table}

\begin{figure}
\centering
\includegraphics[width=3.3in, keepaspectratio=true]{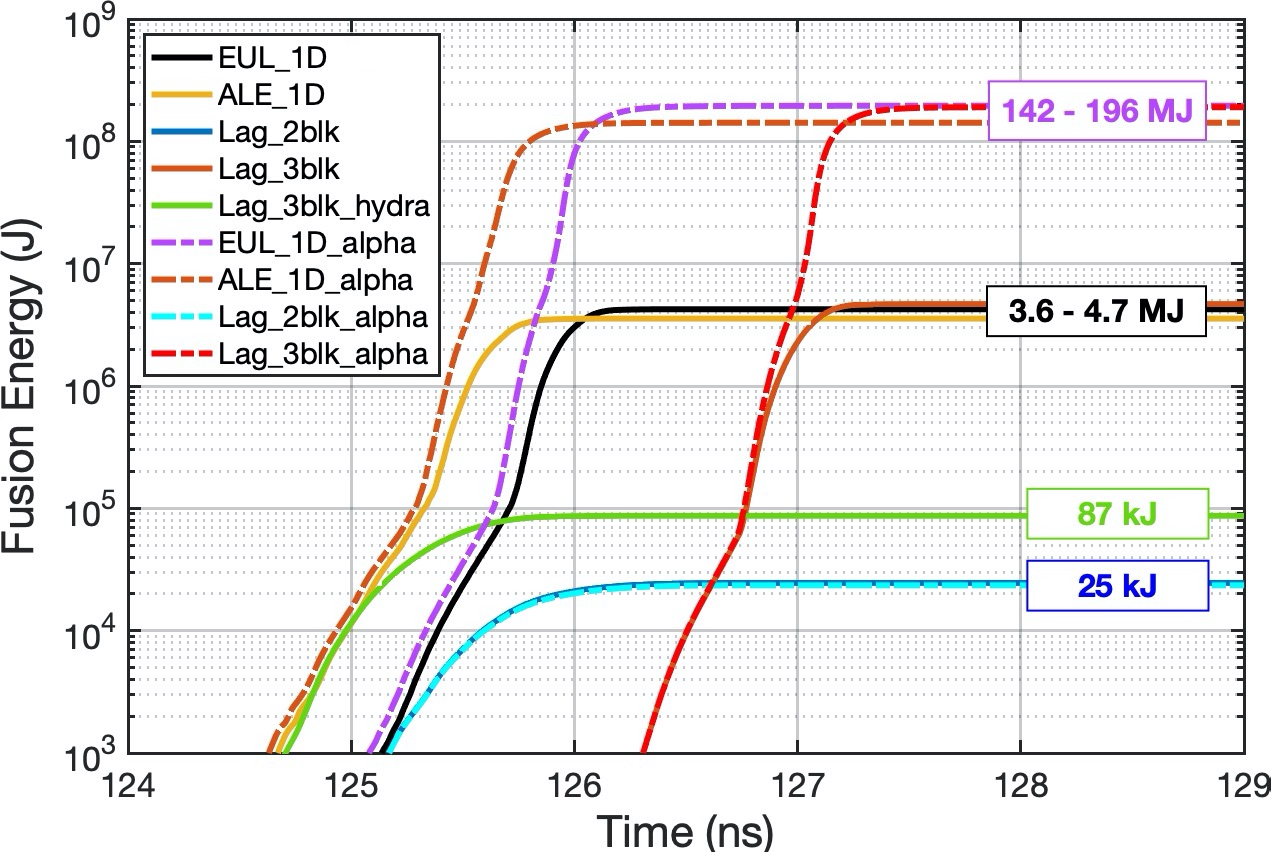}
\caption{ Fusion energy  from   1-D MACH2  simulations, with and without alpha heating. The Lagrangian simulation with a \it hydra \rm suffix is using three blocks,  fixed vacuum boundary and input current profile from the Lindemuth et al. publication. }
\label{fig:6}
\end{figure}

The  pinch stagnation times and the total fusion energy  for the various 1-D  models are summarized in Table I. 
Three models with no $\alpha$-heating   ($Lag\_3blk$, ALE\_1D and EUL\_1D) consistently  calculate about  4 MJ; their prediction is  about 40-50 times higher when  $\alpha$-heating is included. Remarkably, the $Lag\_2blk$  and $Lag\_3blk\_hydra$ models calculate only 30 kJ  and 90kJ;  $\alpha$-heating  can not change these numbers. Visual representation of these results is shown in Fig.6, which also illustrates how the ~1.5ns spread in stagnation times.

A closer look at Fig.3(d)  reveals that from 122 ns to 126 ns the  $Lag\_3blk$ mass averaged fuel temperature grows slower than the corresponding temperatures in the Raven and $Lag\_2blk$
cases.  It reached 1 keV at   t=125 ns, and then in the next 1 ns  it peaked at   2.8 keV. This leads to substantially higher thermal pressure close to the stagnation time $T_s$, as illustrated in Fig.7 where $P_{tot} = P_i + P_e$ contour plots are centered around $T_s$  for each individual simulation. The  EUL\_1D and  $Lag\_3blk$ contour plots look quite similar, with a white central region of high pressure $P_{tot} > 40 GBar$; the ALE\_1D contour plot (not shown) also looks like these two.

The triangular white spaces outside of the liner/vacuum boundary in the two and three block Lagrangian simulations  are due to the shrinkage of the computational domain as the plasma compresses. The other two simulations (EUL\_1D and $Lag\_3blk\_hydra$) have fixed size  domains extending to R=3.1mm.

The small grid size ($\Delta R=1.6\mu m$) in the EUL\_1D simulation allowed for clear resolution of about dozen plasma sound wave fronts, before they are "lost" in the high pressure central region where there is only one color (white). By calculating the speed  of propagation of a wave  front next to the liner/target interface,  these tilted lines indicating regions of higher and lower pressure can easily be identified as compressional sound waves. For example,  one of the middle striations covered 0.051 mm in 0.12 ns, i.e. the velocity is 42.5 cm/$\mu$s, which is  the sound velocity $C_s=(\gamma kT_e/m_i)^{1/2}$ that MACH2 calculates for the target region about 0.5 ns before stagnation.

The magnetic field, a few ns   before the stagnation time, is heavily compressed in the liner, but its pressure at the interface is at least an order of magnitude lower then the thermal pressure, so it can not explain the rise of the mass averaged $T_i$ from 1 keV to 2.8 keV.

\begin{figure}
\centering
\includegraphics[width=3.5in,keepaspectratio=true]{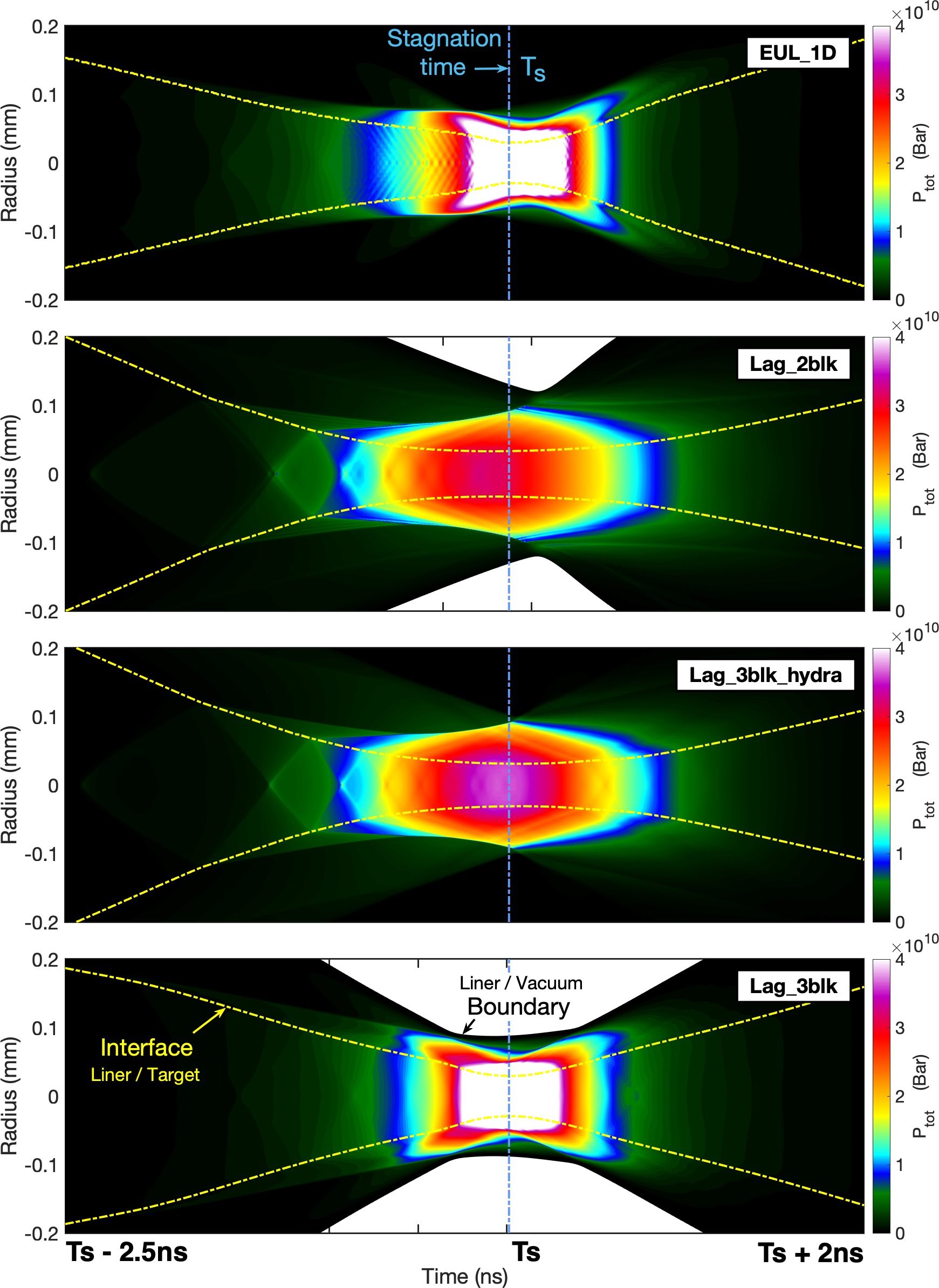}
\caption{ Contour plots of the total plasma thermal pressure   $P_{tot} = P_i + P_e$ for four 1-D MACH2 models.
 The time axis is centered around the stagnation time for each model.}
\label{fig:7}
\end{figure}

\begin{figure*}
\centering
\includegraphics[width=7.in,keepaspectratio=true]{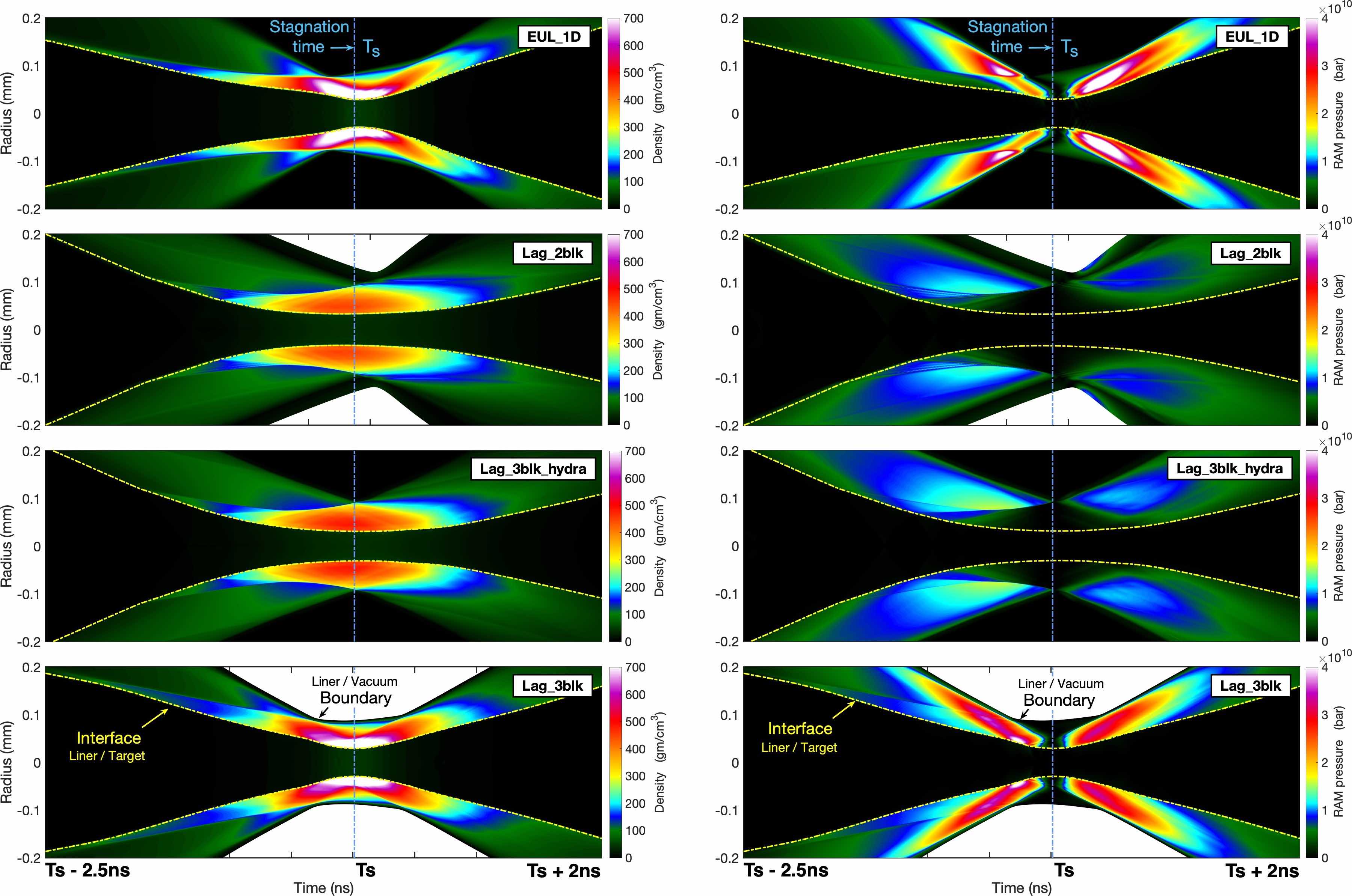}
\caption{Left column: plasma mass density $\rho$ contour plots  for four different 1-D MACH2  SZP models without $\alpha$-heating.   Notice that the EUL\_1D and $Lag\_3blk$ models, about 0.5ns before and after the stagnation time, have significantly larger mass concentration at the liner/target interface. Right column: corresponding ram  pressure   $P_{RAM} = \rho v^2$. The stronger ram pressure  at the liner/target interface during the last ~1.5 ns of the adiabatic compression is responsible for the much higher fusion energy production in the EUL\_1D and $Lag\_3blk$ models.
 The time axis is centered around the stagnation time for each model.}
\label{fig:8}
\end{figure*}

Optically thick radiative (OTR) shocks are of great interest in astrophysics and have been extensively studied both theoretically and experimentally$^{42-44}$.
The Z-machine plasma near stagnation is certainly optically thick.  The   main idea that we want to borrow from the OTR research is that magnetosonic shocks can transport mass to the shock front, creating sharp  profiles with  a density several times higher than the downstream values. A more in-depth discussion on this topic involving Hugoniot equations is beyond the scope of this paper.

Plasma mass density $\rho$ contour plots  for the same set of four 1-D MACH2 simulations from Fig.7 are shown on the left side of Fig.8.  Liner mass accumulation is clearly visible starting $\sim$2 ns before the stagnation time, and it is stronger by a factor of 2-3 for the EUL\_1D and $Lag\_3blk$ models (and the ALE\_1D model, which is not shown) compared to the  $Lag\_blk$ and  $Lag\_blk\_hydra$ models. The liner mass density around $T_s$ is up to  two orders of magnitude  higher than the solid silver density $\rho_{Ag}$ = 10.5 $gm/cm^3$.

The corresponding ram pressure  $P_{RAM} = \rho v^2$ contour plots are shown on the right side of Fig.8. They confirm the several times higher ram pressure for the EUL\_1D and $Lag\_3blk$ models just before stagnation. The higher pressure, through the adiabatic work done on the fuel column, results in a higher fuel thermal energy which can explain why these two models have 2-3 times higher mass averaged fuel ion temperature, compared to the $Lag\_blk$ and  $Lag\_blk\_hydra$ models. 

If magneto-sonic waves are  responsible for piling liner mass at the interface then there must be a variation in the $B_\theta$ magnetic field strength between the models with high and low fusion yield.  In Fig.9  $B_\theta$ profiles  are shown at stagnation time $T_s$ and at four previous nearby times for the same set of four 1-D MACH2 models shown in Fig.7 and Fig. 8.
The magnetic field at the liner/target interface is indeed about 2 times stronger for the  EUL\_1D and $Lag\_3blk$ models. 

The profiles for the $Lag\_3blk\_hydra$ model are particularly instructive. One can visualize the three profiles at  $T_s -2, T_s -1.5,T_s -1 ns$  being similar to the  corresponding profiles for the  EUL\_1D  model, except that their  tops are clipped.  The clipping is more pronounced for the last two profiles, at   $T_s -0.5$ and  $T_s$, and their shape is quite different from the shape of the corresponding profiles for the  EUL\_1D  model. A check of the computational grid for the $Lag\_3blk\_hydra$ model shows that the clipping of the $B_\theta$ field maxima is due to loss of resolution;  there is only one single cell covering the region where $B_\theta$ peaks, resulting in a flat line. 

\begin{figure}
\centering
\includegraphics[width=3.2in, keepaspectratio=false]{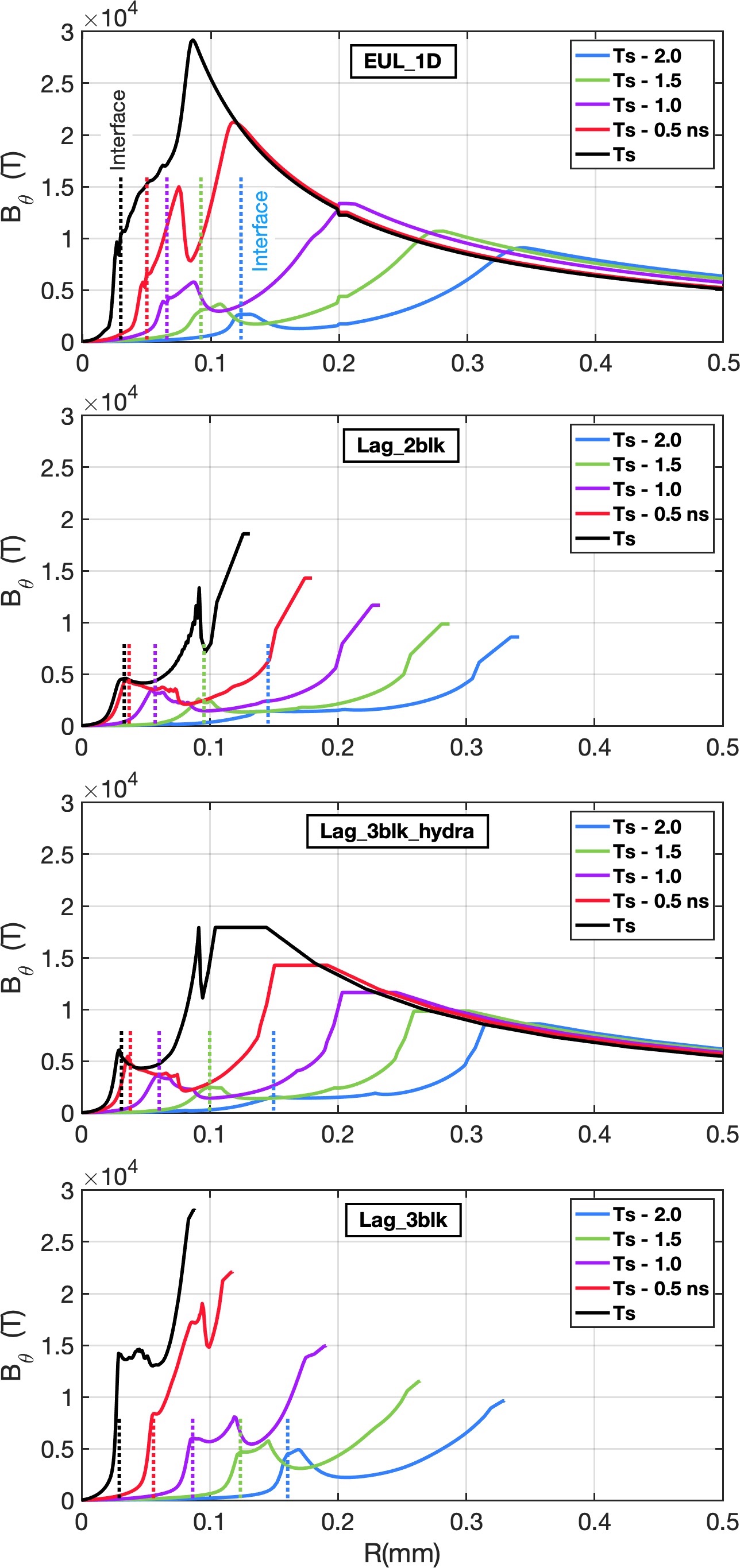}
\caption{ Radial $B_\theta$ profiles  from 1-D  MACH2 calculations using Eulerian, two block pure Lagrangian,  three blocks Lagrangian with fixed vacuum boundary and  three block pure Lagrangian  formalism. The color coded dotted lines indicated the location of the  liner/target interface at a particular time. }
\label{fig:9}
\end{figure}

\begin{figure}
\centering
\includegraphics[width=3.2in, keepaspectratio=true]{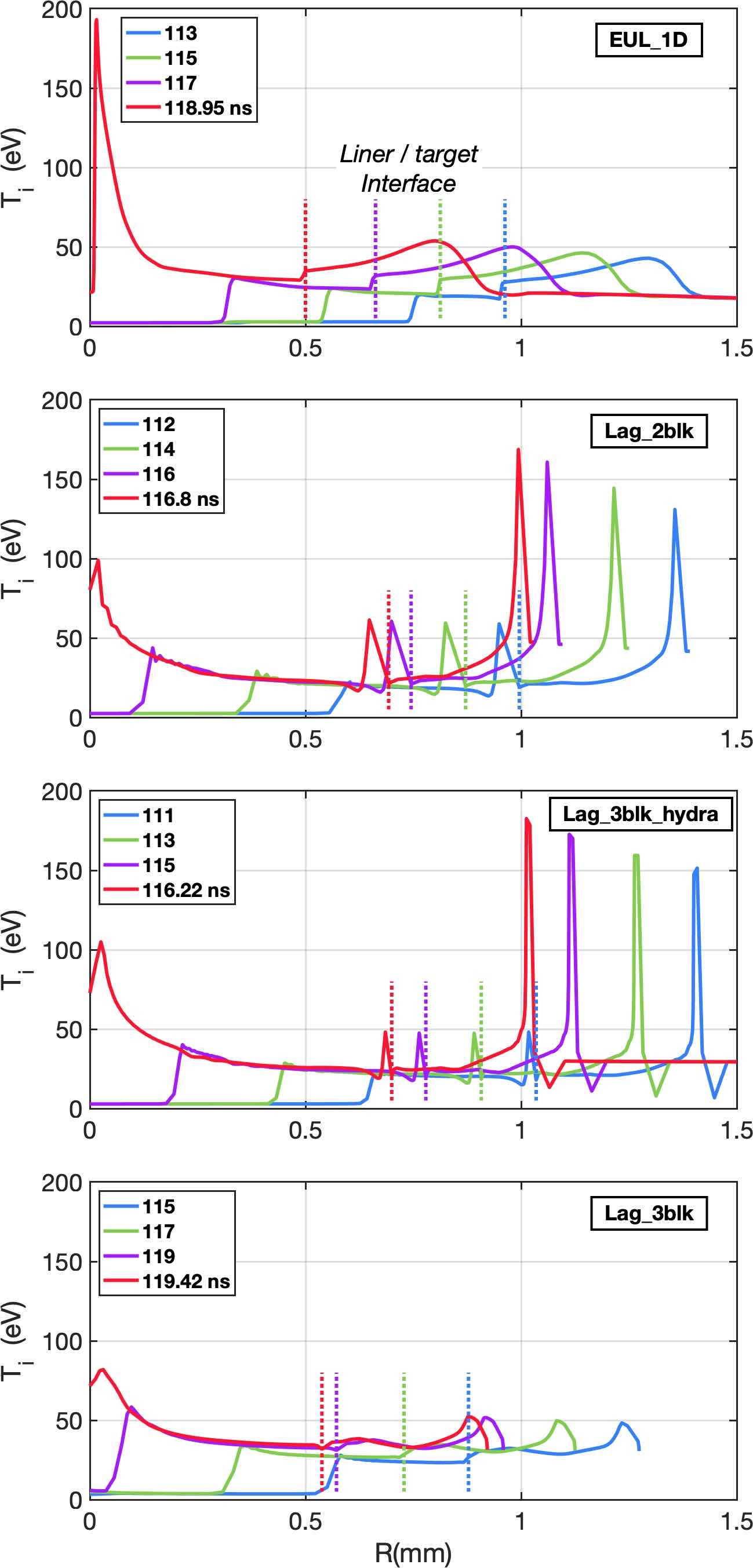}
\caption{Radial $T_i$ profiles  from 1-D  MACH2 calculations using Eulerian, two block pure Lagrangian,  three blocks Lagrangian with fixed vacuum boundary and  three block pure Lagrangian  formalism. The color coded dotted lines indicated the location of the  liner/target interface at a particular time.}
\label{fig:10}
\end{figure}

The ion temperature profiles for four representative times, the last being the instant when the shock reaches the axis,  for the  MACH2 models from Fig.9 are shown in  Fig.10.
They reveal another problem with the   $Lag\_2blk$ and $Lag\_3blk\_hydra$ models: There are sharp peaks at the liner/vacuum boundary and the liner/target interface.  Again, this problem can be traced to single cells covering regions of rapid variable change.  For example, the  $Lag\_3blk$ model has 16 dedicated vacuum cells, and the magnetic field diffusing  from the outermost liner region inside can be more properly calculated. However, proper calculation requires taking into account the $1/R$ dependance of   $B_\theta$ in the vacuum region and  can  be done only with the EUL\_1D or ALE\_1D models.
Improperly calculated $B_\theta$ profile leads to artificially induced very high $J_z$ current  over a few adjacent grid cells,  which then ohmically heats the plasma and produces those $T_i$ profile  spikes.  In the next computational  step, when the diffusion of the $B_\theta$ field is calculated from the plasma conditions in the previous time step, its profile will be affected by the slower magnetic field diffusion through the hot plasma regions; ultimately leading to different  $B_\theta$ profiles and different magnetosonic shock dynamics$^{45}$ compared to those in the EUL\_1D or ALE\_1D models.

At the end of this section we  briefly  touch upon shocks in plasmas, about which there is vast literature that can not be reviewed here. In general, large amplitude waves can propagate at speeds larger than the speed of sound $C_s$. These waves steepen during propagation and the steepening process can be balanced by dispersion and diffusion. If the steepening is balanced by dispersion then this class of waves are called solitary waves which propagate as an isolated finite-amplitude disturbance in plasmas. On the other hand, if steepening is balanced by diffusion it can form a thin layer called shock which then  propagates through the system. During propagation,  shocks separate regions of different density and temperature, and  the shock front exhibits steep gradient in plasma pressure.  Computer simulations must carefully address these gradients,  otherwise the  shocks  role might be lost or underestimated.  
Shock waves are continuously produced  during the SZP implosion,  as long as the liner plasma remains cold and the sound speed $C_s$  is  lower than the implosion speed $V_r$. 
Profiles of the radial compression $V_r$,   sound $C_s$, and  Alfvein velocity $V_A$ (Fig.15) confirm that the liner implodes super-Alfvenically, and the target implodes supersonically.

\section{2-D Simulations of the Staged Z-pinch}

Plasma instabilities are present in any Z-pinch,  therefore the computational grid should cover  two dimensions (R-Z) for a more realistic representation of the SZP dynamics. 
For our Eulerian and ALE   2-D models  the corresponding 1-D grids were extended axially with 64 vertical cells ($\Delta Z=234\mu m$). A 3-D MHD code would bring further refinement to the simulation results, but we do not have access to such a code.  

The 2-D simulations were driven by currents from a circuit model in analogous way, like the Lagrangian simulations presented in Section 2. These currents are compared with the current from Lindemuth et al in Fig.11. The time evolution of the (a) interface radius, (b) interface radial velocity, and (c) mass-averaged fuel temperature are presented in Fig.12. 
The  comparison is very much like the comparison shown in Fig.3. As expected, the peak values of the mass averaged temperatures are somewhat lower:  2.7 keV and 40 keV, depending whether  $\alpha$-particle heating is included or not.

\begin{figure}
\centering
\includegraphics[width=3.3in, keepaspectratio=true]{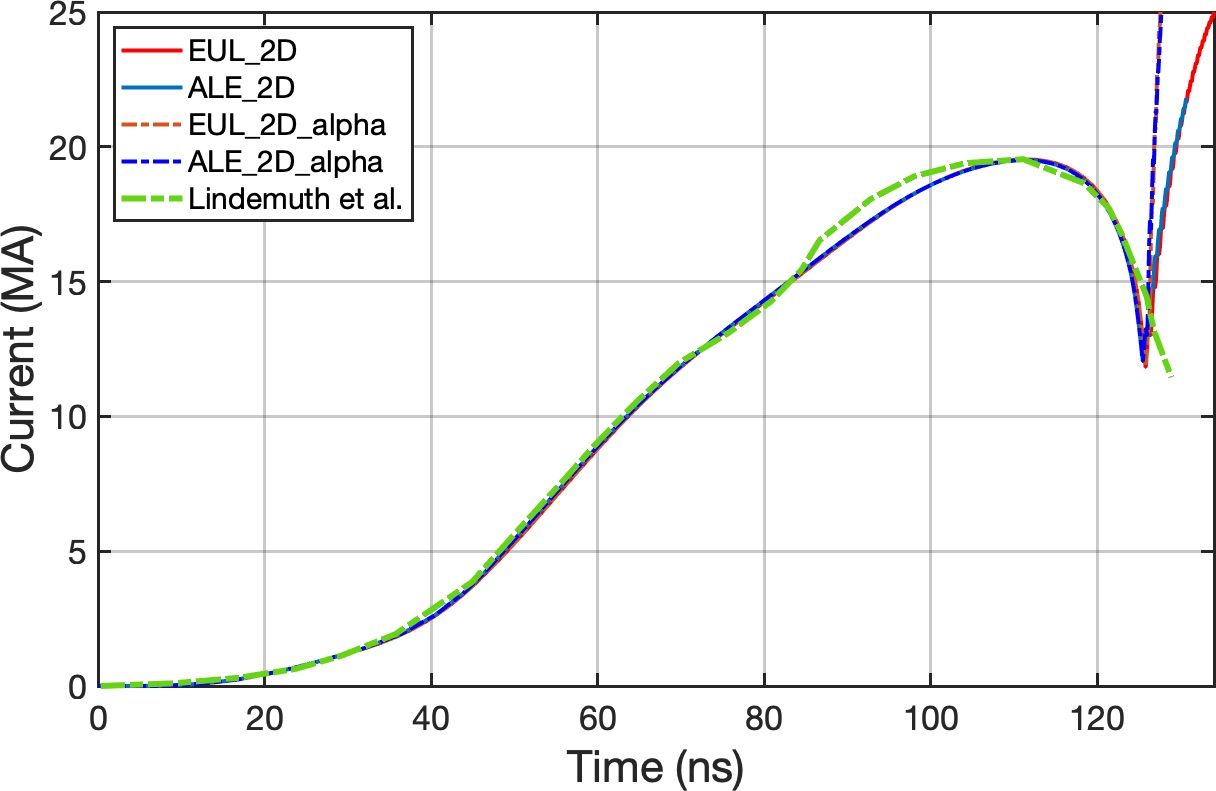}
\caption{Current profiles  from circuit equations driving  2-D Eulerian and ALE  SZP simulations, with and without alpha particle heating.  The dot-dash green current profile is from  the Lindemuth et al. publication$^1$.}
\label{fig:11}
\end{figure}
\begin{figure}
\centering
\includegraphics[width=3.3in, keepaspectratio=true]{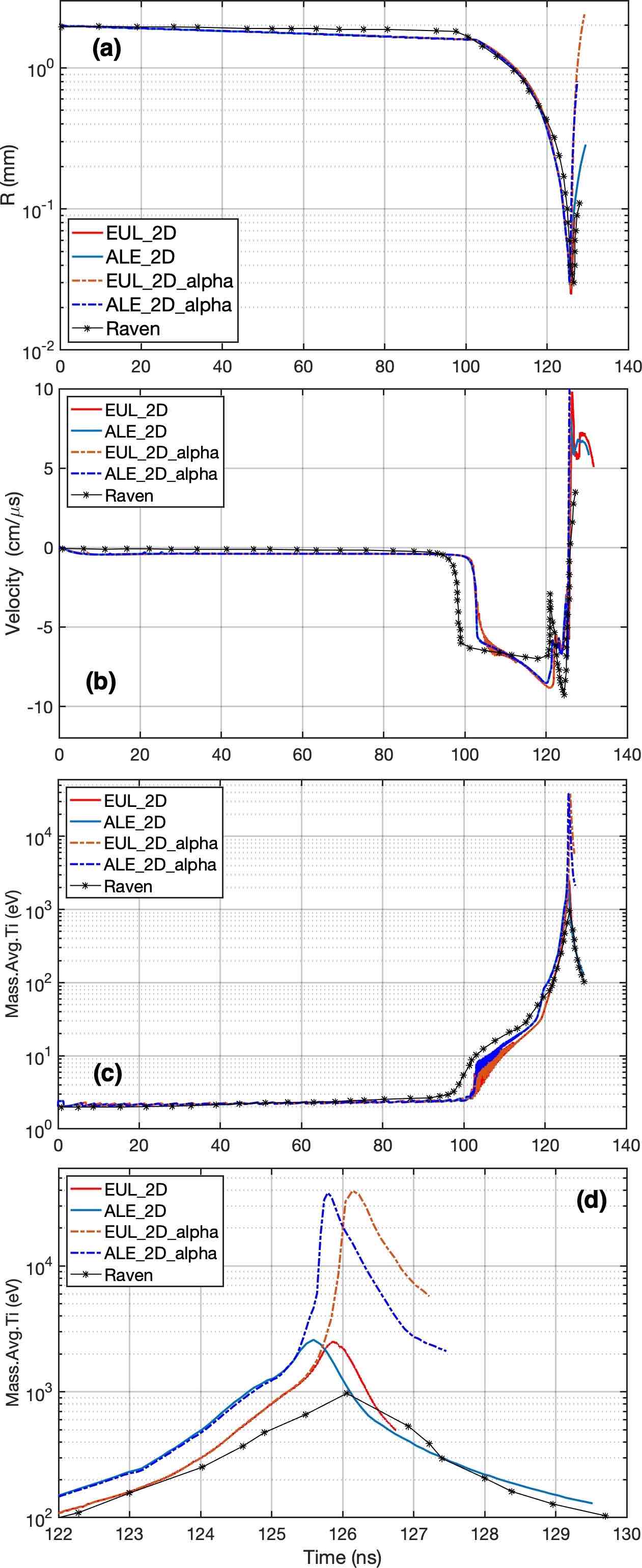}
\caption{Comparison of  2-D Eulerian and ALE  simulations, with and without alpha heating, with results from the Raven code:  (a) radius, (b) implosion velocity  and (c)  mass-averaged fuel temperature, all calculated at the liner/target interface. The large temperature increase when $\alpha$-particle heating is included  is highlighted in (d). }
\label{fig:12}
\end{figure}

The effectiveness of shock heating in raising the target adiabat can be seen by plotting the average target temperature versus target convergence ratio (CR).  For a lossless, cylindrical adiabatic compression of a monatomic ideal gas, the temperature increases as CR$^{4/3}$.  Shock heating or ohmic heating can cause the target to heat super-adiabatically, whereas radiative and conductive losses can lead to sub-adiabatic heating. A typical trajectory of the mass averaged target ion temperature vs. CR is shown in Figure 13. Most of the shock heating occurs during the initial acceleration of the target and concludes when the shock front reaches the axis, at CR$\sim$5.

\begin{figure}
\centering
\includegraphics[width=3.3in, keepaspectratio=true]{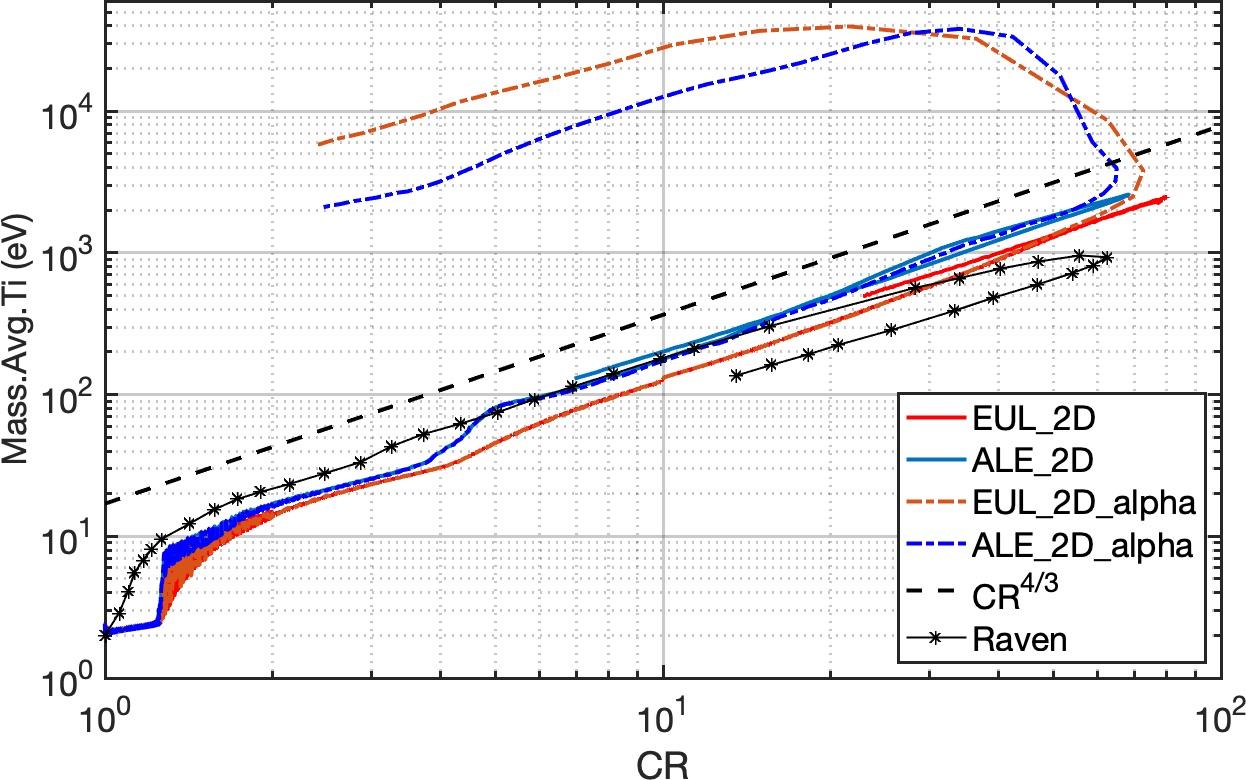}
\caption{ Comparison of  the mass-averaged fuel temperature, calculated at the liner/target interface, vs. compression ratio for 2-D   Eulerian and ALE simulations, with and without alpha heating. The Raven code compression curve is provided for reference, as well as the $\rm CR^{4/3}$ line. Curve sections that raise faster than this line indicate shock heating. }
\label{fig:13}
\end{figure}
\begin{figure}
\centering
\includegraphics[width=3.3in, keepaspectratio=true]{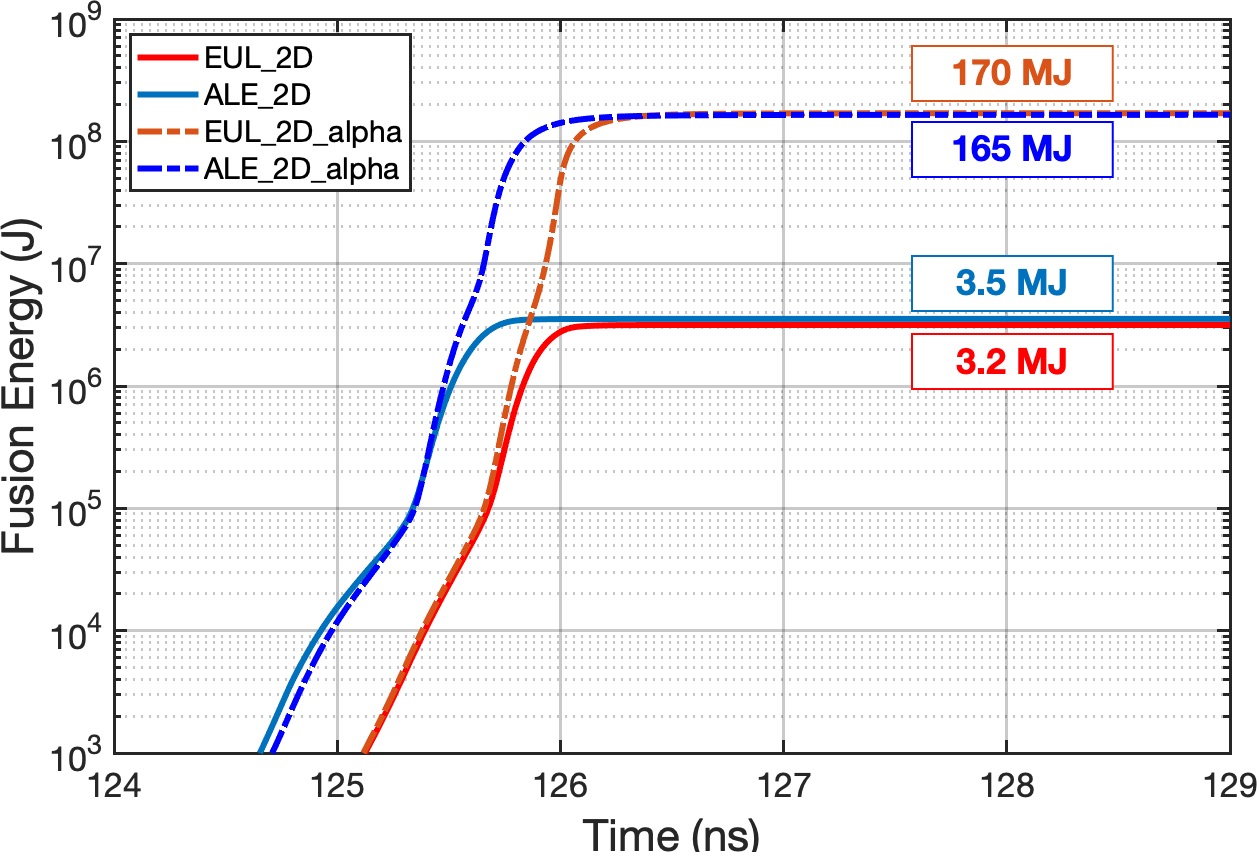}
\caption{Fusion energy  from   2-D MACH2  simulations, with and without alpha heating, using  Eulerian and ALE models.}
\label{fig:14}
\end{figure}

The fusion energy production is shown in Fig.14. It is slightly lower than for the corresponding 1-D models without $\alpha$-heating:  3.2  vs.  4.6 MJ (EUL), and  3.5  vs. 3.6 MJ (ALE). Similarly,  the Eulerian 2-D model with  $\alpha$-heating predicts lower total fusion energy: 170  vs. 196 MJ. However, the prediction for the ALE 2-D model  with  $\alpha$-heating is higher:  165 vs. 142 MJ, close to the 2-D Eulerian prediction, which highlights the importance of proper grid size selection and the somewhat arbitrary grid readjustments in each MACH2 computational block and time instant, when the ALE method is used.

\begin{figure}
\centering
\includegraphics[width=3.2in, keepaspectratio=true]{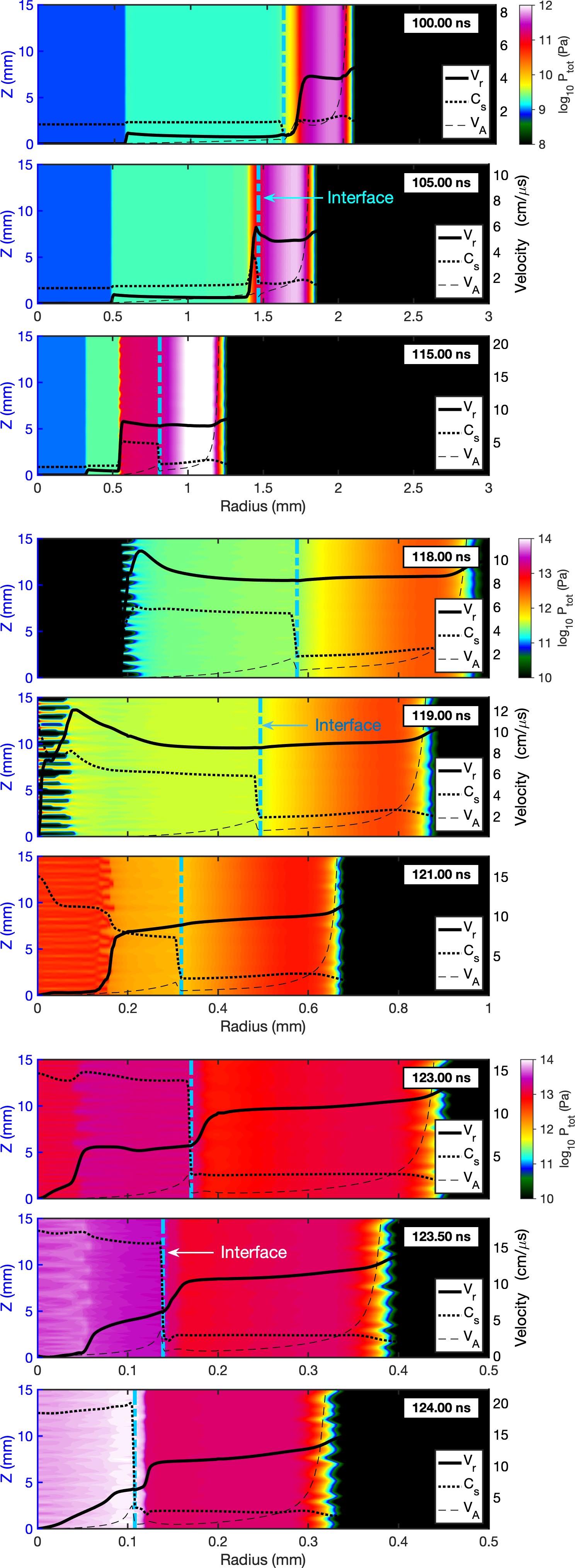}
\caption{Countour plots of the total plasma pressure from a 2-D  Eulerian  calculation, with superimposed  implosion speed $V_r$, sound  speed $C_S$, and Alfvein speed $V_A$, clarifying the  shock wave propagation through the plasma.}
\label{fig:15}
\end{figure}

The Staged Z-pinch shocks are examined in more detail in Fig.15,  where total plasma pressure ($P_{tot} = P_i+P_e$) contour plots at 9 different times within $\sim$25 ns of the pinch stagnation are presented. Profiles of the radial plasma implosion velocity $V_r$, the Alfven velocity $V_A$, and the sound velocity $C_s$ are shown as well. These  velocity profiles confirm  that the liner implodes super-Alfvenically, and the target implodes supersonically. The SZP compression progression is clearly visible in each set of three contour plots (notice the radial scale change in each set), and it picks up pace as the pinch is closer to stagnation. After stagnation (not shown), the pinch radial momentum reverses sign  and the plasma expands rapidly. 

The top three panels of Fig.15 show $P_{tot}$  contour plots at 100, 105, and 115 ns.  At 100 ns, the shock, which is generated in the liner, has still not reached the interface. At 105 ns, it just crossed the interface, and  at 115 ns it propagates to R=0.5 cm, which is well into the target region. Both liner and target plasma keep imploding and  accelerating. The three middle panels   show  the contours at 118, 119, and 121 ns. At 118 ns, the shock front begins to exhibit  Richtmyer-Meshkov type instability$^{46,47}$, and by 119 ns, the unstable shock front reaches  the axis (note the skewed aspect ratio of the plots: The vertical dimension is compressed up to 30 times with respect to the radial dimension).

We run MACH2 on a modern Linux workstation built around the  Intel Xeon E5 2690 v4 processor (2.6GHz,  14 cores and 28 threads). However, MACH2 is a single thread code and it takes about 100 hours to complete the 2-D ALE and EUL models with $\alpha$-particle heating. The unexpectedly high  wall time expended on the ALE\_2D model, in spite of the 2-3 coarser grid than the EUL\_2D model, was perhaps due to the readjustments of the computational grid at each time step of the calculation.

\section{Alpha particle heating and pinch radiation considerations}
Now we focus on the crucial role of $\alpha$-particle heating by closely examining our  2-D Eulerian simulations. The  MACH2 $\alpha$-heating implementation assumes 100\% $\alpha$-particle energy deposition into the D-T fuel plasma;  this assumption will be revisited shortly. 

\begin{figure}
\centering
\includegraphics[width=3.3in,keepaspectratio=true]{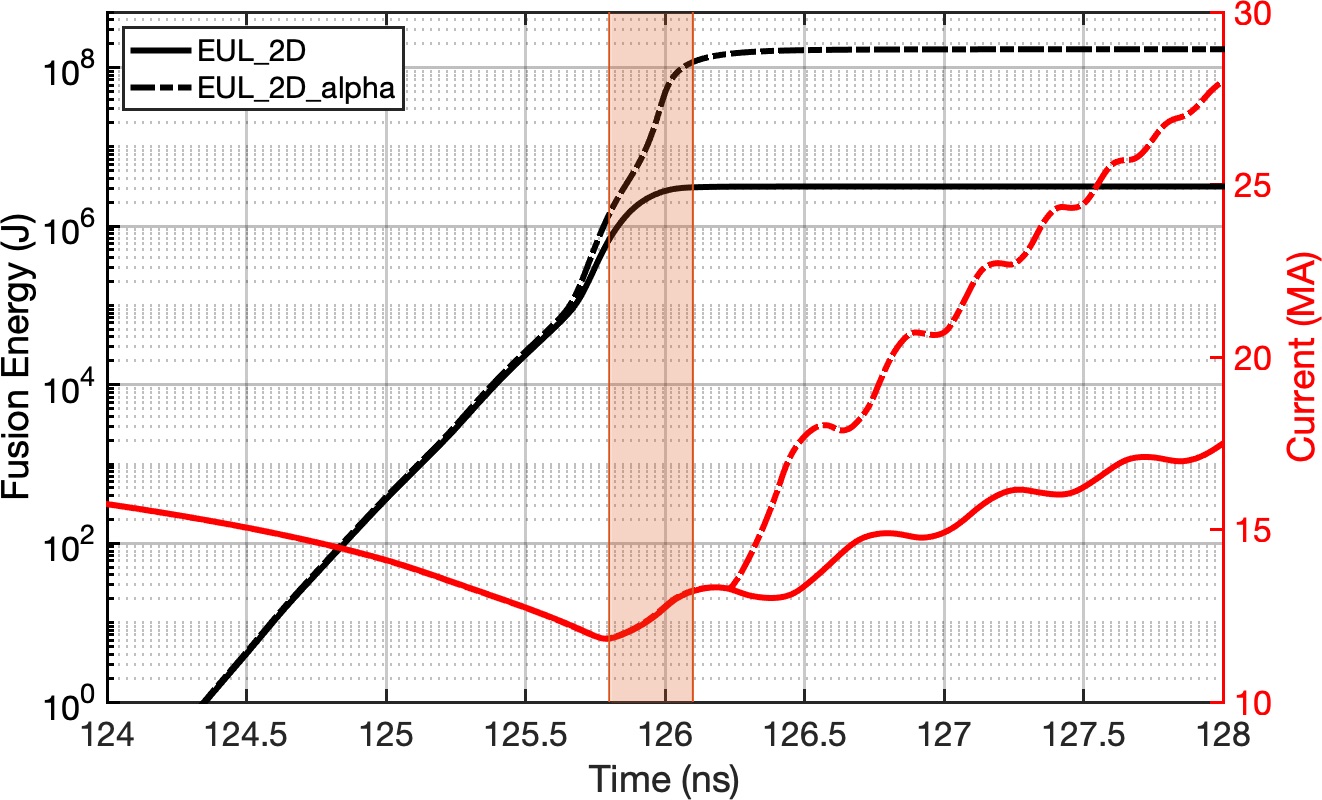}
\caption{ Alpha particle heating illustration: in just 0.3 ns (shaded region) the fusion energy production increases from 1.4 MJ to 118 MJ. The pinch  stagnation time for the model with alpha heating is 125.82 ns, and 125.84 ns  if the alpha heating is turned off. }
\label{fig:16}
\end{figure}

Figure 16 illustrates the dramatic effect of $\alpha$-heating by comparing the fusion energy calculated in 2-D Eulerian models with and without such heating; it takes only 0.3 ns to increase the fusion energy from 1.4 MJ to 118 MJ !  The plasma current does not change until 126.24 ns (when 150 MJ are already produced, i.e. 88\% of the total 170 MJ), and then it starts growing much faster than the current in the model without    $\alpha$-heating.

Figure 17 clarifies the internal plasma dynamics during the $\alpha$-heating phase by showing contour plots of the ion density $N_i$ and ion temperature $T_i$ at four times: 125.9, 125.94, 126.07 and 126.24 ns, when the fusion energy is 5.5, 10, 100 and 150 MJ, respectively. The azimuthal magnetic field $B_\theta$ is shown as well, in appropriate colors to provide contrast with the contour plots; outside of the liner region i.e. in the vacuum region $B_\theta$ decreases as 1/R. The radial $B_r$ and axial $B_z$ components of the magnetic field are zero everywhere.

\begin{figure} 
\centering
\includegraphics[width=3.3in, keepaspectratio=true]{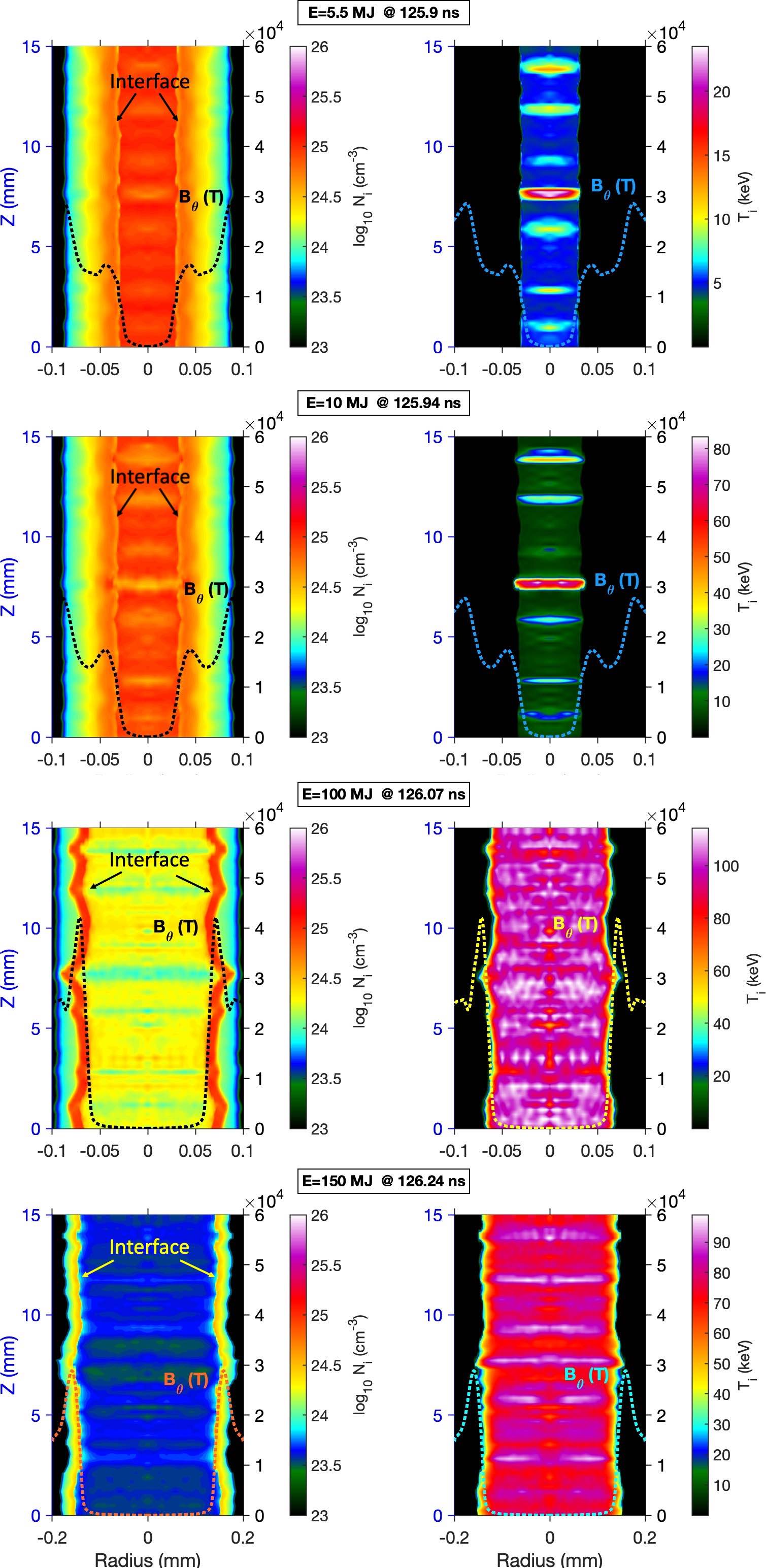}
\caption{ Ion density and temperature contour plots for four $\alpha$-heating time instances which demonstrate the dramatic rise in fusion energy production from 5.5MJ to 150 MJ, in just 0.34 ns. 
The superimposed $B_\theta$ magnetic field profiles  are  sharply peaked  near the liner/target interface where $B_\theta > 10^4 $ T  provides good $\alpha$-particle confinement. Data is from the Eulerian 2-D simulation. }
\label{fig:17}
\end{figure}

The fuel ion density is roughly between $10^{24}$ and $10^{25}cm^{-3}$ for the first three selected times.  For the last, the target radius had approximately doubled (notice the radial scale change) and the fuel density deceased several times, as 1/$R^2$. The ion temperature at 125.9 ns is about 5 keV for most of the fuel plasma, with a single narrow region reaching 20-25 keV. After 0.17ns, by which time 100 MJ of fusion energy is produced, most of the fuel volume has temperatures of about 100 keV. In the next 0.17 ns, while the plasma column keeps expanding, additional 50 MJ of energy are produced, in a slightly colder plasma, with several times lower density.

The $B_\theta$ magnetic field profile has very high value at the liner/target interface; approximately  between $10^4$ and $3\times 10^4$ T.   This field has been continuously growing  by  diffusing through the liner, and by liner compression (Fig.9), until the stagnation time.  After stagnation,  $\alpha$-heating builds enormous pressure which can not instantaneously reverse the radial motion of the large liner mass. This pressure rapidly compresses the magnetic field at the interface to levels  above $10^4$ T, and that is the essence of the magneto-inertial confinement mechanism in the Staged Z-pinch: The extremely strong $B_\theta$ field helps with the  $\alpha$-particle confinement  and reduces the thermal energy exchange between the very hot fuel and cold liner plasma. 

There are two questions to be answered regarding $\alpha$-heating: Are the $\alpha$-particles confined and do they have enough time to deposit their energy to the background plasma? 
The second question is easily answered by comparing the  $\alpha$-particle energy slowing down time$^{48}$ on electrons, $\tau_{\alpha | e, slw}$ , within the $\alpha$-heating time window $\Delta t \approx 0.35ns$. At 125.9 ns, $N_e\approx10^{25}cm^{-3}, T_e \approx 5 keV$ and   $\tau_{\alpha | e, slw} \approx 10^{-3} ns$, and since $\tau_{\alpha | e, slw} \sim T_e ^{3/2} /N_e$, this slowing down time will not change dramatically. It takes $t=3\times\tau_{\alpha | e, slw}$ to deposit 95\% of the 3.5 MeV $\alpha$-particles energy, therefore alphas have plenty of time to transfer their energy to the fuel plasma.

The 3.5 MeV $\alpha$-particle  larmor radius in a $B_\theta=10^4 T$ magnetic field is $\rho_L = 27 \mu m $, which is comparable to the radius of the target region during the $\alpha$-heating period.  Its velocity is  $v_\alpha= 1.3\times 10^7 m/s$,  so it covers a 0.44 mm path length in $\Delta t =0.34 ns$. Taking into account the very fast slowing down on electrons, this path length will quickly shorten, and thus only a small fraction of alphas may be lost by reaching the vertical boundaries of the simulation domain (Z=0 and Z=15mm).  

Finally, let's consider the effects of possible fuel depletion due to the high fusion rate during the $\alpha$-heating phase. The central fuel volume is V=2.83 $cm^3$ and contains 6.6$\times10^{21}$ of deuterium, and the same number of tritium nuclei; their fusion produces a total of 6.1$\times10^{19}$ 14 MeV neutrons and same number of 3.5 MeV alphas, i.e. 170 MJ. Therefore, only about 1\%  of the fuel  is consumed ("burn-up" fraction), which can be ignored, since MACH2 does not have a burn-up calculation which depletes the fuel density as the high fusion rate proceeds in the target.

\begin{table*}
\caption{Radiation loss and $PdV$ heating estimates for the 1-D Lagrangian 2-block simulation using the same methods as in Lindemuth's  2017 paper$^{49}$. Modification of the radiation loss estimates to account for nonzero $T_r$ is shown to lower loss estimates below $PdV$ heating except near stagnation, for t = 125.8 ns. Note that $n_0=2.36\times10^{21}$ cm$^{-3}$ and $r_0=3$ mm.} 
\vspace{3mm}
\label{table3}
\centering
\def\arraystretch{1.5}
\begin{tabular}{|c |c |c |c |c |c|}   \hline
Time (ns) & ~~~~~~122  ~~~~~~& ~~~~~~ 123  ~~~~~~& ~~~~~~ 124  ~~~~~~&  ~~~~~~125  ~~~~~~& ~~~~~~ 125.8  ~~~~~~\\ \hline \hline
$r_\mathrm{target}$ (m) & 3.18$\times10^{-4}$ & 2.29$\times10^{-4}$ & 1.21$\times10^{-4}$ & 4.73$\times10^{-5}$ & 3.08$\times10^{-5}$ \\ \hline 
$v_\mathrm{target}$ (cm/$\mu$s) & 4.43 & 10.46 & 10.70 & 5.49 & 0.15  \\    \hline
$\langle n_\mathrm{target} \rangle$ (cm$^{-3}$) & 9.31$\times10^{22}$ & 1.81$\times10^{23}$ & 6.48$\times10^{23}$ & 4.22$\times10^{24}$ & 9.93$\times10^{24}$ \\ \hline
$\langle T_e\rangle$ (eV) & 70.94 & 118.74 & 266.53 & 767.21 & 978.62 \\ \hline
$\langle T_r\rangle$ (eV) & 70.88 & 118.52 & 263.21& 753.34 & 974.62 \\ \hline
$\bf P_{brems} (W/m^3)$ & 1.23$\times10^{21}$ & 6.01$\times10^{21}$ & 1.16$\times10^{23}$ & 8.32$\times10^{24}$ & 5.20$\times10^{25}$ \\ \hline
$(T_e^4-T_r^4)/T_e^4$ & 3.38$\times10^{-3}$ & 7.39$\times10^{-3}$ & 4.89$\times10^{-2}$ & 7.04$\times10^{-2}$ & 1.65$\times10^{-3}$ \\ \hline
$\bf P_{brems,corrected}(W/m^3)$ & 4.34$\times10^{18}$ & 4.42$\times10^{19}$ & 5.67$\times10^{21}$ & 5.86$\times10^{23}$ & 8.46$\times10^{23}$ \\ \hline
$\bf P_{PdV} (W/m^3)$ & 5.89$\times10^{20}$ & 6.29$\times10^{21}$ & 9.82$\times10^{22}$ & 2.41$\times10^{24}$ & 3.02$\times10^{23}$ \\ \hline
\end{tabular}
\end{table*}

\begin{table*}
\caption{Radiation loss and $PdV$ heating estimates for the 1-D Eulerian simulation with $\alpha$-particle heating, using the same methods as in Lindemuth's  2017 paper$^{49}$. Modification of the radiation loss estimates to account for nonzero $T_r$ is shown to lower loss estimates below $PdV$ heating. Note that $n_0=2.36\times10^{21}$ cm$^{-3}$ and $r_0=3$ mm. The calculated $\alpha$-particle heating is also included, and shown to exceed radiation losses near peak compression, indicating ignition.}
\vspace{3mm}
\label{table4}
\centering

\def\arraystretch{1.5}
\begin{tabular}{|c|c|c|c|c|c|}
\hline
Time (ns) & ~~~~~~ 122   ~~~~~~& ~~~~~~ 123  ~~~~~~& ~~~~~~ 124  ~~~~~~& ~~~~~~ 125  ~~~~~~& ~~~~~~ 125.8  ~~~~~~\\ \hline \hline
$r_\mathrm{target}$ (m) & 2.23$\times10^{-4}$ & 1.70$\times10^{-4}$ & 1.07$\times10^{-4}$ & 5.55$\times10^{-5}$ & 2.89$\times10^{-5}$ \\ 
\hline
$v_\mathrm{target}$ (cm/$\mu$s) & 6.29 & 6.26 & 6.47 & 3.23 & 2.98 \\ \hline
$\langle n_\mathrm{target} \rangle$ (cm$^{-3}$) &1.90$\times10^{23}$ & 3.28$\times10^{23}$ & 8.17$\times10^{23}$ & 3.06$\times10^{24}$ & 1.13$\times10^{25}$ \\ \hline
$\langle T_e \rangle$ (eV) & 118.72 & 174.04 & 328.32 & 868.14 & 3991.35 \\ \hline
$\langle T_r \rangle$ (eV) & 116.18 & 173.51 & 327.90 & 865.06 & 3787.51 \\ \hline
$\bf P_{brems} (W/m^3)$ & 6.66$\times10^{21}$ & 2.40$\times10^{22}$ & 2.04$\times10^{23}$ & 4.67$\times10^{24}$ & 1.36$\times10^{26}$ \\ \hline
$(T_e^4-T_r^4)/T_e^4$ & 8.29$\times10^{-2}$ & 1.21$\times10^{-2}$ & 5.11$\times10^{-3}$ & 1.41$\times10^{-2}$ & 1.89$\times10^{-1}$ \\ \hline
$\bf P_{brems,corrected} (W/m^3)$ & 5.53$\times10^{20}$ & 2.87$\times10^{20}$ & 1.05$\times10^{21}$ & 6.59$\times10^{22}$ & 2.57$\times10^{25}$ \\ \hline
$\bf P_{PdV} (W/m^3)$ & 4.08$\times10^{21}$ & 1.35$\times10^{22}$ & 1.03$\times10^{23}$ & 9.89$\times10^{23}$ & 2.96$\times10^{25}$ \\ \hline
$P_\alpha$ (W/m$^3$) & 3.22$\times10^{11}$ & 8.71$\times10^{13}$ & 2.32$\times10^{17}$ & 5.52$\times10^{21}$ & 1.81$\times10^{26}$ \\ \hline
\end{tabular}
\end{table*}

Simple estimates of radiation and thermal conduction losses and $PdV$ heating rates can be used as another check of a  MHD code simulation results.  Estimates in Lindemuth's 2017 paper$^{49}$ (Table III) show that radiation losses are greater than the $PdV$ heating, but they assume  radiation temperature $T_r=0$. In reality, the radiation temperature inside the target is nonzero because the liner is optically thick. In SZP2 it is assumed that $T_r$ equilibrates with the electron temperature $T_e$ instantaneously, i.e. that the radiation losses are zero. The radiation loss estimates by Lindemuth for SZP2 therefore cannot be used to invalidate the SZP2 results. 

Because $T_r = T_e$ is a strong assumption, the results presented in this paper use a radiation diffusion model that allows $T_r$ to evolve dynamically. As shown in our Table II, $T_r = T_e$ is in closer agreement with the conditions near stagnation than $T_r = 0$. Estimates of the bremsstrahlung radiation should therefore be reduced by the factor $(T_e^4-T_r^4)/T_e^4$, which brings the estimated radiation loss rates below the $PdV$ heating rates until stagnation. These results clearly  indicate the possibility of ignition, which occurs when the  $\alpha$-particle heating exceeds the plasma energy losses. 

In Table III we  compare the $\alpha$-particle heating rates with the heating and loss estimates for the  EUL\_1D\_alpha simulation.  Again,  the $PdV$ heating is larger than the  corrected bremsstrahlung estimates, and at peak compression the $\alpha$-particle heating exceeds even the $T_r=0$ radiation loss estimate, confirming the ignition predicted by MACH2.

\section{Conclusions}

We used the MACH2 code in various Lagrangian, Eulerian and arbitrary Lagrangian Eulerian (ALE) modes to simulate the Staged Z-pinch dynamics where a thin silver liner implodes onto  D-T fuel in the Sandia Laboratories Z-machine.  By using a two block Lagrangian model, and  a  three block Lagrangian model with a fixed boundary, we reproduced the results of Lindemuth et al.\cite{Lindemuth:2018}  and thus verified the code. We then pointed out that the likely problem with the Hydra, Raven and MHRDR results is the incorrect treatment of the liner/vacuum boundary. Insufficient computational grid resolution in this critical region   leads to calculating  incorrect  $B_\theta$  profiles,  and they affect the strength of the magneto-sonic shocks which are responsible for piling mass at the liner/target interface. With proper liner/vacuum boundary treatment,  MACH2 indicates that there is extra  liner mass accumulated at the interface, which increases the ram pressure in  the final implosion stages; the associated $PdV$ work adiabatically transfers the liner kinetic energy into fuel thermal energy, raises its mass-averaged temperature from 1 kV to $\sim$2.5 keV, and sets the stage for successful $\alpha$-particle heating.

 We  believe that,  in spite of the limitations of the MACH2 code, these results merit careful review with codes and material tables not available in the public domain. If confirmed, only experiments on the Z-machine can validate them and hopefully surpass  break-even fusion energy production. 

\vfill

\begin{acknowledgments}
Funding for this work was provided by ARPA-E, grant No. DE-AR0000569, and by Strong Atomics.
We acknowledge useful discussions with Dr. M. H. Frese and Dr. J. Narkis.
\end{acknowledgments}

\end{document}